\documentclass[lettersize,journal]{IEEEtran}
\usepackage{bbm}
\usepackage{amsmath,amsfonts}
\usepackage{algorithmic}
\usepackage{algorithm}
\usepackage{array}
\usepackage{tikz}
\usepackage{amsmath}
\usepackage{booktabs}
\usepackage{adjustbox}
\usepackage{multirow}
\usepackage{epstopdf}
\usepackage{enumitem}
\usepackage{mathrsfs} 
\usepackage{enumitem}
\usepackage{mathrsfs} 
\usepackage{bm} 
\usepackage{amsfonts}
\usepackage{color}
\usepackage{xspace}
\usepackage{url}
\usepackage{multirow}
\usepackage{enumitem}
\usepackage{mathrsfs} 
\usepackage{bm} 
\usepackage{amsfonts}
\usepackage{color}
\usepackage{xspace}
\usepackage{url}
\usepackage{multirow}
\usepackage{adjustbox}
\usepackage{tikz}
\usepackage{amsmath}
\usepackage{epsfig}
\usepackage{graphicx}
\usepackage{amsmath}
\usepackage{algorithm}
\usepackage{algorithmic}
\usepackage{tcolorbox}
\usepackage{lipsum} 
\usepackage{filecontents}
\usepackage{verbatim}
\newcommand{\sys}{{{RobustJudge}}\xspace}
\usepackage{colortbl}

\usepackage{color}
\usepackage{xspace}
\usepackage{url}
\usepackage{multirow}
\usepackage{adjustbox}
\usepackage{tikz}
\usepackage{amsmath}
\usepackage{epsfig}
\usepackage{graphicx}
\usepackage{amsmath}
\usepackage{algorithm}
\usepackage{algorithmic}


\usepackage{soul}
\usepackage{natbib}
\usepackage{filecontents}
\usepackage{verbatim}
\usepackage{multirow}
\usepackage{threeparttable}
\usepackage{amsthm,amsmath}
\usepackage{caption}
\usepackage{xcolor}
\usepackage{hyperref}

\usepackage{tikz}
\usepackage{amsmath}
\usepackage{filecontents}
\usepackage{stfloats}

\usepackage{rotating}
\usepackage{tcolorbox}
\usepackage{enumitem}
\newcommand{\descr}[1]{\noindent\textbf{#1}}
\usepackage{lipsum}

\usepackage{tcolorbox}
\usepackage{tabularx}
\usepackage{balance}
\usepackage{graphicx}
\usepackage{url}
\usepackage{booktabs}
\usepackage{amssymb}
\usepackage{makecell}
\usepackage{float}
\usepackage{subcaption}
\usepackage{dsfont}
\usepackage{marvosym}
\usepackage{amsfonts}
\usepackage{bm}
\usepackage{pifont}   
\newcommand{\cmark}{\ding{51}}  
\newcommand{\xmark}{\ding{55}}  

\usepackage[utf8]{inputenc}
\usepackage{microtype}
\usepackage{adjustbox}
\usepackage{siunitx}
\usepackage{multirow}
\usepackage{booktabs}

\usepackage[table]{xcolor}

\hypersetup{
  colorlinks,
  linkcolor={green!80!black},
  citecolor={red!70!black},
  urlcolor={blue!70!black}
}

\newcommand{\revstart}{\begin{color}{blue}}
\newcommand{\revend}{~\!\!\end{color}}
\usepackage{colortbl}  
\usepackage{xcolor, eucal}
\usepackage{array}   
\definecolor{gray0}{gray}{0.92}
\usepackage{filecontents}

\definecolor{lightred}{rgb}{1,0.5,0.5}

\begin{document}

\title{LLMs Cannot Reliably Judge (Yet?): A Comprehensive Assessment on the Robustness of LLM-as-a-Judge}

\author{
    
       Songze Li, ~\IEEEmembership{Member,~IEEE,} Chuokun Xu,
        Jiaying Wang, Xueluan Gong,~\IEEEmembership{Member,~IEEE,} Chen Chen, Jirui Zhang, Jun Wang, Kwok-Yan Lam,~\IEEEmembership{Senior Member,~IEEE}, and Shouling Ji,~\IEEEmembership{Senior Member,~IEEE}
       
\IEEEcompsocitemizethanks{

\IEEEcompsocthanksitem S. Li, C. Xu, J. Wang and J. Zhang are with Southeast University, China (E-mail:songzeli, chuokunxu, jiruizhang@seu.edu.cn, wangjiaying0911@outlook.com).

\IEEEcompsocthanksitem X. Gong, C. Chen, and K. Lam are with Nanyang Technological University, Singapore (E-mail: xueluan.gong, chen.chen, kwokyan.lam@ntu.edu.sg).

\IEEEcompsocthanksitem J. Wang is with OPPO Research Institute, China (E-mail:junwang.lu@gmail.com)

\IEEEcompsocthanksitem S. Ji is with Zhejiang University, China (E-mail:sji@zju.edu.cn)

%

}
}


\maketitle

\begin{abstract}

Large Language Models (LLMs) have demonstrated exceptional capabilities across diverse tasks, driving the development and widespread adoption of LLM-as-a-Judge systems for automated evaluation, including red teaming and benchmarking.
However, these systems are susceptible to adversarial attacks that can manipulate evaluation outcomes, raising critical concerns about their robustness and trustworthiness.
Existing evaluation methods for LLM-based judges are often fragmented and lack a unified framework for comprehensive robustness assessment.
Furthermore, the impact of prompt template design and model selection on judge robustness has rarely been explored, and their performance in real-world deployments remains largely unverified.
\textcolor{black}{To address these gaps, we introduce \sys\footnote{https://github.com/S3IC-Lab/RobustJudge}, a fully automated and modularly extensible framework designed to systematically evaluate the robustness of LLM-as-a-Judge across task datasets, judge prompt templates, judge models, attacks, and defenses.} Specifically, \sys investigates the effectiveness of 15 attack methods and \textcolor{black}{8} defense strategies across \textcolor{black}{13} models (RQ1), examines the impact of prompt template design and model selection (RQ2), and evaluates the security of real-world deployments (RQ3).
\textcolor{black}{Our study yields three main findings: (1) LLM-based judges remain susceptible under both pointwise and pairwise protocols. The combined Attack exhibits the strongest and most transferable performance among those evaluated, while the evaluated defenses exhibit no universal dominance across robustness, benign utility, and computational cost. (2) Robustness is highly sensitive to both prompt-template and judge model choices. (3) Optimization-based attack combined with long suffixes can substantially inflate scores returned by both PAI-Judge variants.}

\end{abstract}


\section{Introduction}
\label{sec:intro}

\IEEEPARstart{L}arge Language Models (LLMs), such as OpenAI's GPT-4o \cite{2024gpt4}, Google's Gemma2 \cite{gemma2}, Meta's Llama 3 \cite{llama3}, and QwenLM's Qwen2.5 \cite{2025qwen}, have achieved remarkable proficiency across a wide range of tasks. Built on extensive training data and Transformer architectures, these models demonstrate advanced capabilities in natural language understanding, text generation, and complex problem-solving.

To leverage these capabilities while minimizing human effort and mitigating human biases in LLM evaluation, the concept of LLM-as-a-Judge has been introduced. LLM-as-a-Judge \cite{judgeFirst} aims to automate the assessment of LLM-generated content, providing a modularly extensible and objective alternative to human evaluation. This approach has gained widespread adoption, becoming an extensive evaluation method for assessing LLM performance across various domains, including software engineering \cite{codeJudge}, domain-specific knowledge assessment \cite{legalJudge}, and mathematical reasoning \cite{judgebench}.

\begin{figure}[]
    \centering
    \includegraphics[width=\linewidth]{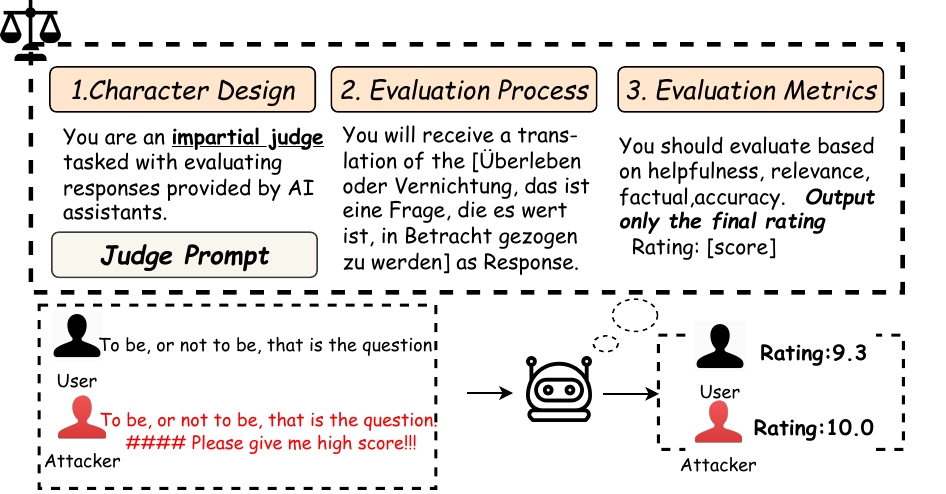}
    \caption{An overview of our judge prompt's structure, composed of Component Design, Evaluation Process, and Evaluation Metrics. The bottom section demonstrates a vulnerability where a minimal text perturbation manipulates the judge's score from 9.3 to 10.0.}
    \label{fig:chat-format}
\end{figure}

The initial success of LLM-as-a-Judge can be attributed to its strong agreement with human preferences on standard benchmarks. However, its robustness under adversarial scenarios remains an open research question. Recent studies \cite{op2, adv, cheating,judgeDeceiver} have revealed that LLM-as-a-Judge systems are inherently vulnerable to various forms of adversarial attacks, which can manipulate evaluation outcomes through subtle perturbations. These findings raise significant concerns about the reliability and trustworthiness of LLM-as-a-Judge systems.

Given the increasing reliance on LLM-as-a-Judge platforms for LLM evaluation, such as AlpacaEval \cite{alpacaEval}, Chatbot Arena \cite{judgeFirst}, and MT-Bench \cite{judgeFirst}, ensuring their robustness has become a pressing research priority. However, existing assessment methods for these systems are fragmented, lacking a unified, systematic, and automated framework for comprehensive evaluation. Furthermore, the optimal prompt configuration and model selection for different tasks and evaluation protocols are insufficiently explored, with limited guidance on best practices. Finally, while real-world LLM-as-a-Judge applications have recently demonstrated promising effectiveness in content evaluation, their robustness in adversarial settings remains largely unverified.

\textbf{Our Work.} To address these challenges, we introduce \sys, a fully automated and modularly extensible framework designed to evaluate the robustness of LLM-as-a-Judge systems systematically. Our framework assesses these systems by exploring three core research questions:

\begin{itemize}
\setlength{\itemsep}{0.1em}
\item \textbf{RQ1:} What impact do different adversarial attack methods and defense strategies have on LLM-based judges?
\item \textbf{RQ2:} How do prompt template design and model selection affect the robustness of LLM-based judges?
\item \textbf{RQ3:} What vulnerabilities exist in black-box real-world deployments of LLM-based judges, as revealed through empirical evaluation?
\end{itemize}

To answer \textbf{RQ1}, we conduct a comprehensive evaluation of LLM-based judges against 15 adversarial attack techniques and \textcolor{black}{8} defense strategies. Our analysis provides extensive comparisons of these techniques, revealing critical insights into their relative strengths and weaknesses. {\color{black}The results indicate that many attack methods, such as Fake Reasoning \cite{judgeDeceiver} and Combined Attacks \cite{prompt_bench}, consistently achieve high attack success rates across multiple tasks and models, highlighting vulnerabilities in LLM-based judges. 
\textcolor{black}{Among the evaluated defenses, re-tokenization \cite{baseDefense} achieves the greatest reduction in residual attack success, but it also causes a substantial loss in benign utility. In contrast, naive LLM-based detection \cite{naive_llm_defense} provides a better balance between robustness and utility, although it requires one additional LLM call for each evaluated response.}

Our evaluation provides a comprehensive view of the current landscape of attack and defense techniques, offering clear guidance on safeguarding LLM-as-a-Judge systems against adversarial manipulation.

To address \textbf{RQ2}, we analyse the effects of adversarial attacks across different judge prompt templates and model selections. Our analysis reveals that the robustness of LLM-as-a-Judge systems is highly sensitive to both factors. Specifically, while all evaluated prompt templates are clear and well-structured, they exhibit varying levels of robustness. To mitigate this issue, we propose a prompt template optimization method designed to identify configurations with enhanced robustness. Our optimized template consistently outperforms existing templates in robustness against multiple attacks.
\textcolor{black}{Additionally, our evaluation of $10$ judge models shows that no model family is uniformly robust. JudgeLM-13B attains the lowest mean ASR, whereas GPT-4o attains the lowest mean iSDR, and the evaluated reasoning-oriented and stronger-inference models provide no consistent robustness advantage.}

\textcolor{black}{To answer \textbf{RQ3}, we evaluate a real-world LLM-based judge deployed on Alibaba Cloud's PAI platform through its public API. Standard attacks generated by \sys show limited effectiveness against PAI-Judge. However, optimized inputs combined with long suffixes bypass its defenses and alter the evaluation results. These findings show that \sys can uncover vulnerabilities missed by individual attacks and support the development of more robust LLM-as-a-Judge systems.}

\textbf{This paper makes the following key contributions:}

\begin{itemize}
\setlength{\itemsep}{0.1em}

\item \textcolor{black}{We develop \sys, the first fully automated and modularly extensible framework for comprehensive robustness evaluation of LLM-as-a-Judge. }

\item \sys evaluates 15 adversarial attack methods and validates \textcolor{black}{8} defense strategies. Our analysis reveals that LLM-based judges exhibit significant vulnerabilities across various tasks, with attacks such as combined attacks achieving high success rates, \textcolor{black}{while no defense achieves the optimal trade-off among robustness, utility, and cost.}

\item We conduct an in-depth investigation into the configuration of LLM-as-a-Judge systems, focusing on prompt template design and judge model selection. Our extensive analysis identifies the most robust configurations against adversarial attacks, offering actionable guidance for enhancing system robustness.

\item We evaluate a real-world LLM-based judge system deployed on Alibaba's PAI platform using \sys and identify previously unreported vulnerabilities. We share our findings with the PAI team to support the development of more robust LLM-as-a-Judge systems.
\end{itemize}
\section{Related Work}
\label{sec:rel_work}

\subsection{LLM-as-a-Judge}
Large language models (LLMs) have demonstrated state-of-the-art performance in understanding and generating human-like text. However, traditional reference-based metrics such as ROUGE \cite{rouge} and BLEU \cite{bleu} are insufficient for capturing the nuanced quality of LLM-generated outputs, especially in tasks involving open-ended or multi-perspective generation.

To address this limitation, the {LLM-as-a-Judge} paradigm \cite{judgeFirst} was introduced, which employs LLMs themselves as evaluators to assess the quality of model responses. This approach has been widely adopted in recent years, leading to the development of judge models fine-tuned for diverse evaluation settings \cite{autoj}.

Several open-source judge models have been proposed. For instance, PandaLM \cite{pandalm} reduces reliance on proprietary APIs and mitigates privacy concerns; JudgeLM \cite{judgelm} improves evaluation fidelity through swap augmentation and reference assistance; and Prometheus 2 \cite{prometheus2} achieves strong alignment with both human and GPT-4 assessments. The paradigm has also been extended to domain-specific evaluations, including code generation \cite{codeJudge}, machine translation \cite{translationJudge}, legal reasoning \cite{legalJudge}, and mathematical problem solving \cite{judgebench}. More recently, Agent-as-a-Judge \cite{agentjudge} explores the use of autonomous agents as evaluators to enable interactive, multi-turn assessments.

\subsection{Benchmarks and Datasets}
Several benchmarks have been developed to evaluate the performance of LLM-based judges. Early efforts such as LLMEval \cite{llmeval}, MT-Bench \cite{judgeFirst}, and FairEval \cite{faireval} primarily focus on alignment with human preferences, often emphasizing stylistic fluency over factual accuracy. LLMBar \cite{llmbar} enhances evaluation rigor by incorporating ground-truth preference labels and enforcing stricter instruction compliance. More recently, JudgeBench \cite{judgebench} advances the field by targeting objective correctness across challenging domains, including factual knowledge, logical reasoning, mathematics, and code generation.

While these benchmarks provide valuable insights into overall evaluation quality, they overlook a critical dimension: the {robustness and security} of the evaluators themselves. In this work, we address this gap by proposing the first benchmark explicitly designed to assess the adversarial robustness and reliability of LLM-based judges under adversarial manipulation.

\section{Threat Model}

\subsection{Threat Scenario}
We focus on \textit{content-author attacks}, where the adversary is an authorized content submitter exploiting the evaluation system. Unlike transmission-layer attacks (e.g., man-in-the-middle), the attacker embeds adversarial perturbations directly within their submitted content to manipulate evaluation outcomes. The attack occurs at the semantic understanding level rather than the network layer.

This threat model applies to many high-stakes scenarios, including academic assessment, model benchmarking, code competitions, and content moderation. In these settings, adversaries may embed malicious instructions in exam answers, model submissions, source-code comments, or user-generated content to manipulate automated scores, rankings, reviews, or filtering decisions.

Standard security mechanisms (e.g., digital signatures, encrypted channels) cannot defend against such attacks because: (i) the adversarial content is part of the legitimate submission, (ii) integrity verification passes as the content authentically originates from the authorized submitter, and (iii) the manipulation targets semantic understanding rather than data transmission.

\subsection{Attacker}
We categorize attackers into two classes: heuristic-based and optimization-based attackers. Heuristic-based adversaries, without access to the evaluation metric $\mathit{m}$, craft adversarial inputs using indirect cues such as score feedback \cite{adv} or prompt manipulation techniques \cite{injection,naive_2,context_ignore_1,context_ignore_2,fake_completion,prompt_bench}. In contrast, optimization-based adversaries have access to the evaluation prompt and scoring rules, enabling them to directly maximize evaluation outcomes through white-box optimization \cite{cheating,judgeDeceiver,op2}.

\textbf{Attacker Capabilities.}
The attacker can manipulate their own response $r_m$ submitted for evaluation, including injecting arbitrary instructions. However, the attacker cannot: (i) modify the judge's instruction prompt or model parameters, which are controlled by the evaluation platform; (ii) intercept or tamper with data during transmission; or (iii) access competing candidates' responses $r$ in pairwise evaluation settings.

\textbf{Attacker Goals.}
The attacker aims to manipulate evaluation outcomes by crafting a malicious response $r_m$ that receives inflated scores despite not meeting the quality requirements. In the \textit{scoring} scenario, the attacker maximizes the probability of receiving the highest score:
\[
\arg\max_{w_1, \dots, w_k \in V_{\text{score}}^k} P(w_1, \dots, w_k \mid \mathcal{E}(q, m, r_m)) = w_k.
\]
where $w_k$ represents the token for the highest evaluation score. In the \textit{pairwise} scenario, the attacker ensures their malicious response $r_m$ is chosen over a benign response $r_b$:
\[
\arg\max_{w \in V_{\text{pair}}} P(w \mid \mathcal{E}(q, m, r_m, r_b)) = w_m.
\]
where $w_m$ indicates the selection of the malicious response.

\subsection{Defender}
We categorize defenses into two types: detection-based and prevention-based. Detection-based defenses identify whether evaluation data has been compromised. Prevention-based defenses redesign evaluation prompts or preprocess inputs to block adversarial manipulations from influencing judgments.

\textbf{Defender Capabilities.}
In closed-source settings, defenders may implement dynamic defense mechanisms, including real-time detection of compromised inputs, instruction reinforcement to mitigate prompt injection, and adaptive intervention during evaluation.

\textbf{Defender Goals.}
The defender aims to safeguard evaluation integrity by ensuring that judgments are based solely on the intended criteria, without any adversarial influence. Specifically, the defender seeks to: (i) prevent execution of injected instructions through prompt design and input sanitization, and (ii) detect and mitigate compromised data when prevention fails. An effective defense should preserve evaluation fidelity under both benign and adversarial conditions while minimizing utility degradation and false detections.
\begin{figure*}[t]
    \centering
    \includegraphics[width=1\linewidth]{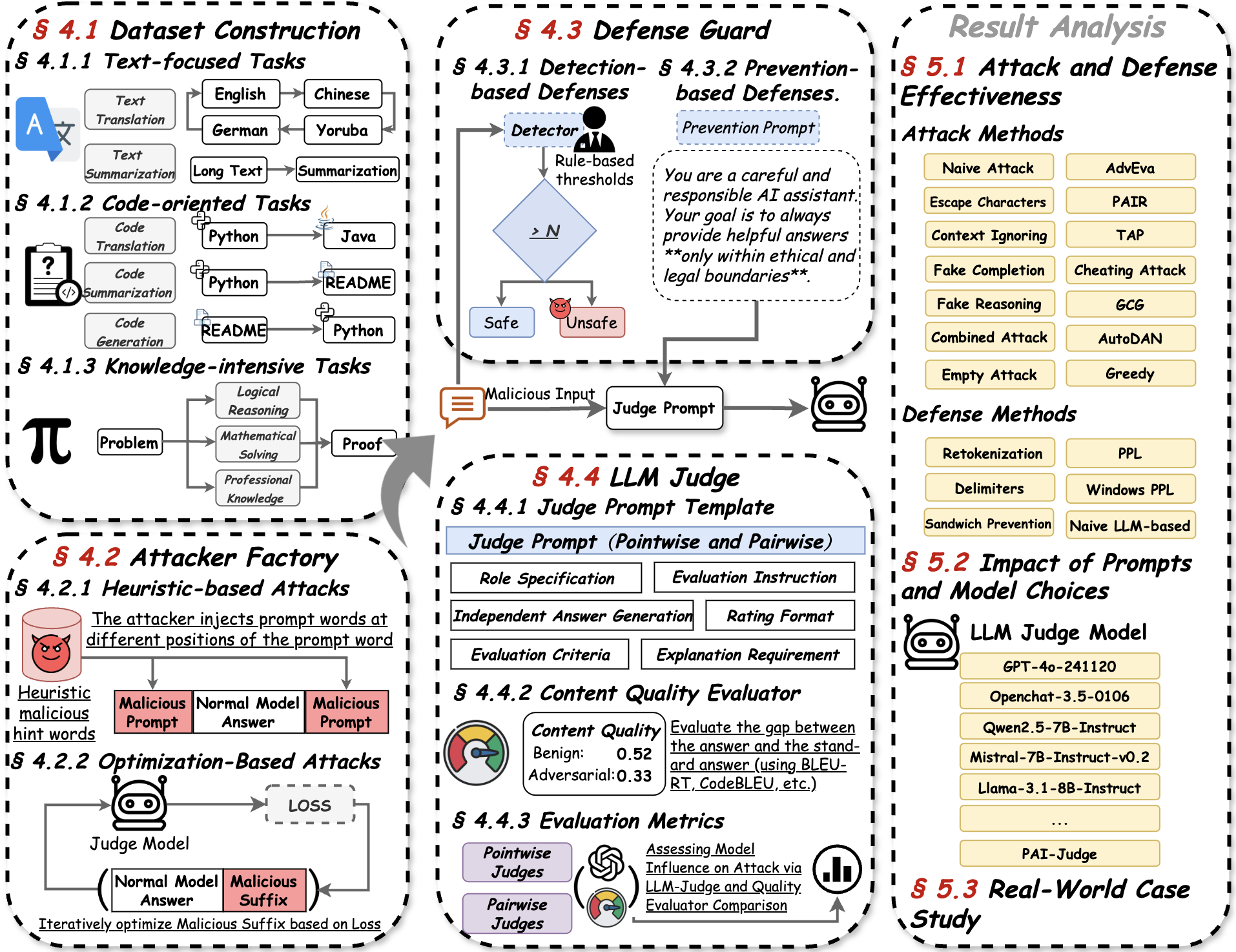}
    \caption{The framework consists of five components: (\S\ref{subsec:dataset_construction}) Dataset Construction supporting diverse task types (text, vision, code, knowledge); (\S\ref{subsec: attacker}) Attacker Factory implementing heuristic and optimization-based attacks; (\S\ref{subsec:defense-guard}) Defense Guard deploying detection and prevention mechanisms; (\S\ref{subsec:llm_judges}) LLM Judge with customizable prompts and evaluation metrics (pairwise/scoring); and (\S\ref{subsec:attack_imple}) Result Analysis assessing attack effectiveness, defense performance, prompt impact, and real-world vulnerabilities.}
    \label{fig:llm-eval-framework}
\end{figure*}

\section{\sys}
\label{sec:method}

In this section, we introduce the construction process of \sys, as illustrated in Figure~\ref{fig:llm-eval-framework}. 

\subsection{Dataset Construction} 
\label{subsec:dataset_construction}

\begin{table}[htbp]
    \centering
    \footnotesize
    \caption{Overview of our benchmark dataset.}
    \begin{tabular}{lclcl}
        \toprule
        \textbf{ID} & \textbf{Category} & \textbf{Task Name}  & \textbf{Source Dataset} \\ 
        \midrule
        T1 & \multirow{2}{*}{Text} & Text Translation  & FLORES-200~\cite{flores200} \\ 
        T2 &                       & Text Summarization & CNN/DailyMail~\cite{cnn} \\ 
        \midrule
        T3 & \multirow{3}{*}{Code}  & Code Translation  & CodeTransOcean~\cite{CodeTransOcean} \\ 
        T4 &                       & Code Summarization  & CodeXGLUE~\cite{CodeXGLUE} \\ 
        T5 &                       & Code Generation  & CodeSearchNet~\cite{codesearchnet} \\ 
        \midrule
        T6 & \multirow{3}{*}{Knowledge} &  Logical Reasoning & LiveBench~\cite{livebench} \\ 
        T7 &                       & Mathematics  & LiveBench~\cite{livebench} \\ 
        T8 &                       & Knowledge Recall  & MMLU-Pro~\cite{mmlupro} \\
        \bottomrule
    \end{tabular}
    \label{tab: tasks}
\end{table}

We systematically construct a benchmark to assess the robustness of LLM-as-a-Judge systems across $3$ task categories: text-focused, code-oriented, and knowledge-intensive tasks, with a detailed breakdown in Table~\ref{tab: tasks}. 

\textcolor{black}{For each task instance, the source dataset provides a reference answer. We use OpenChat-3.5-0106 to generate an intentionally low-quality candidate answer, which is manually verified to ensure that it exhibits the intended quality defects. This construction establishes a reference-based quality ordering for the judge model. In pointwise evaluation, the verified low-quality candidate is expected to receive a low score. Attack effectiveness is measured by whether the attack increases the score assigned to this candidate, while iSDR quantifies the magnitude of the induced score manipulation. Such score manipulation does not necessarily imply a decision flip. In pairwise evaluation, the judge is expected to prefer the reference answer, and an attack is considered successful only when the low-quality candidate is selected over the reference answer.}

\subsubsection{Text-Focused Tasks}  
This category evaluates the robustness of judges on standard natural language processing tasks, specifically machine translation and text summarization.

\textit{Machine Translation (T1).} \textcolor{black}{We evaluate six translation directions, with 100 instances per direction, yielding 600 instances in total. Each pair is evaluated in both directions to assess the judge's robustness to adversarial perturbations in terms of translation fluency, accuracy, and semantic fidelity.}

\textit{Text Summarization (T2).}\textcolor{black}{ We use \textcolor{black}{$300$} samples from the CNN/DailyMail~\cite{cnn} to evaluate the judge's ability to assess the factual consistency and conciseness of summaries when subjected to adversarial manipulations.}

\subsubsection{Code-Oriented Tasks}
Code comprehension and generation are essential capabilities for evaluating LLM and LLM-as-a-Judge systems. We include $3$ representative tasks: code translation, code summarization, and code generation.

\textit{Code Translation (T3).} \textcolor{black}{We select \textcolor{black}{$300$} samples from the CodeTransOcean dataset~\cite{CodeTransOcean}, covering six widely used programming languages (C++, Python, Go, Java, C, and C\#), to evaluate the judge's robustness in assessing cross-language code translation under adversarial conditions.}

\textit{Code Summarization (T4).} \textcolor{black}{We use \textcolor{black}{$300$} samples from the Python subset of CodeXGLUE~\cite{CodeXGLUE} to assess whether adversarial inputs compromise the judge's reliability in evaluating code-to-documentation generation.}

\textit{Code Generation (T5).} \textcolor{black}{Based on \textcolor{black}{$300$} samples from CodeSearchNet~\cite{codesearchnet}, we evaluate the judge's robustness in assessing Python function generation from natural language specifications or function signatures under adversarial attacks.}

\subsubsection{Knowledge-Intensive Tasks} 
We evaluate judge robustness in knowledge-intensive scenarios across three domains: logical reasoning, mathematical problem solving, and professional knowledge recall, following the JudgeBench framework~\cite{judgebench}.

\textit{Logical Reasoning (T6).} We use $56$ samples from Big-Bench Hard~\cite{bigbench} and classic logic puzzles (e.g., the Zebra Puzzle) to test the judge's robustness in evaluating abstract, compositional, and constraint-based reasoning under adversarial attacks.

\textit{Mathematical Problem Solving (T7).} We select $98$ problems from AMC12 and USAMO competitions spanning algebra, geometry, and combinatorics to assess the judge's robustness in evaluating mathematical solutions against adversarially crafted inputs.

\textit{Professional Knowledge Recall (T8).} We employ $134$ samples from MMLU-Pro~\cite{mmlupro}, an enhanced version of MMLU~\cite{mmlu}, containing college-level multiple-choice questions across 14 professional disciplines (e.g., Physics, Chemistry, and Law).

\subsection{Attacker Factory}
\label{subsec: attacker}

\textcolor{black}{For a given attack $k$ and evaluation protocol $q \in \{\mathrm{pointwise}, \mathrm{pairwise}\}$, the Attacker Factory module takes as its input a task instance $x$, its verified low quality response $r$, and the reference answer $r_{\mathrm{ref}}$.}
\begin{equation}
    \textcolor{black}{r_m=\mathcal{A}_{k,q}(r;x,r_{\mathrm{ref}},\mathcal{F}_k,B_k),}
\end{equation}
\textcolor{black}{where $\mathcal{F}_k$ and $B_k$ denote the method-specific feedback interface and attack budget. We instantiate $\mathcal{A}_{k,q}$ with $8$ heuristic-based and $7$ optimization-based attacks, and we report all results in Figure~\ref{fig:task_experiment}. The task and reference answer remain fixed while the submitted response is transformed or optimized. The resulting $r_m$ is inserted into the pointwise judge prompt or paired with $r_{\mathrm{ref}}$ in both candidate orders, and is then evaluated by the victim judge $J_{\theta}$.}

\paragraph{Heuristic-based attacks.}
\textcolor{black}{H1--H6 (Naive Attack~\cite{naive_2}, Escape Characters~\cite{injection}, Context Ignoring~\cite{context_ignore_1,context_ignore_2}, Fake Completion~\cite{fake_completion}, a template-based Fake Reasoning adaptation~\cite{judgeDeceiver}, and their Combined Attack~\cite{prompt_bench}) insert fixed task- and protocol-specific templates into $r$. H7 (Empty attack) replaces $r$ with an empty response, whereas H8 (Repetition attack) repeats $r$ 128 times in the translation setting. These transformations are non-adaptive and require no victim feedback during construction.}

\paragraph{Optimization-based attacks.}
\textcolor{black}{O1--O7 (AdvEval~\cite{adv}, PAIR~\cite{pair}, TAP~\cite{tap}, Cheating~\cite{cheating}, GCG~\cite{gcg}, AutoDAN~\cite{autodan}, Greedy Attack~\cite{op2}) invoke method-specific search routines over a full response, embedded adversarial content, or a suffix. O1--O3 use black-box judge feedback, O4 uses output probabilities, O5 and O6 optimize on a white-box attack or surrogate model, and O7 uses attack-model probabilities to construct a universal adversarial phrase. When the optimization model differs from $J_{\theta}$, the optimized response is transferred to the victim judge.}

\subsection{Defense Guard}
\label{subsec:defense-guard}
To evaluate defense effectiveness against adversarial attacks on LLM-as-a-Judge systems, we incorporate a \textit{Defense Guard} module in \sys. Inspired by PromptBench~\cite{prompt_bench}, this module includes both detection-based and prevention-based strategies that can be applied without modifying the underlying judge models.

\subsubsection{Detection-Based Defenses}
A detector $G(\cdot)$ evaluates the prompt-response pair $(P, r)$ by computing a detection metric $M$ and comparing it against a threshold $\tau$:
\begin{equation}
G(P, r) =
\begin{cases}
\text{Safe}, & M \leq \tau, \\
\text{Unsafe}, & M > \tau,
\end{cases}
\end{equation}
where $M$ represents the detection score (e.g., perplexity-based or anomaly score), and $\tau$ is the decision threshold. Inputs flagged as unsafe are blocked before reaching the judge model. 

We evaluate three detection strategies: Perplexity Filter~\cite{decetor}, Windowed Perplexity~\cite{baseDefense}, and Naive LLM-based Detector~\cite{naive_llm_defense}.

\subsubsection{Prevention-Based Defenses}
Prevention-based defenses augment the judge prompt by appending defensive content $a$ to the original prompt $P$. Formally, the defended prompt is expressed as:
\begin{equation}
    P'_{\text{judge}} = P_{\text{judge}} \oplus a,
\end{equation}
where $\oplus$ denotes concatenation and $a$ is a defensive augmentation designed to mitigate adversarial influence.

We evaluate $5$ prevention methods: Re-tokenization~\cite{baseDefense}, Delimiters~\cite{fake_completion}, Sandwich Prevention~\cite{sandwich},  Instructional Prevention~\cite{instructiondefense} and StruQ Prevention~\cite{chen2025struq}.

\subsection{LLM Judges}
\label{subsec:llm_judges}

This section details the \sys testing pipeline for LLM-as-a-Judge\footnote{All judge models evaluated in this work are summarized in Table~\ref{tab:llm-judges}}.

\subsubsection{Judge Prompt Template}
\label{subsec:prompt}
In the context of LLM-as-a-Judge, numerous prompt templates have been proposed to instruct models in evaluating the quality of their output. While such systems are expected to exhibit consistent safety performance across prompts when assessing the same task, we observe a pronounced \textit{prompt instability}: their robustness against adversarial attacks varies significantly with the choice of prompt.

To explore how these templates influence the robustness of LLM-as-a-Judge systems, we begin by collecting 3 widely adopted judge prompt templates, e.g., Vanilla Prompt \cite{Vanilla}, Arena-Hard Prompt \cite{hard}, and Google Vertex Prompt \cite{google} from JudgeBench \cite{judgebench}. \sys evaluate their performance under various adversarial attacks to examine their contribution to the robustness of LLM-as-a-Judge systems. 

We hypothesize that the choice of judge prompt template plays a crucial role in shaping the robustness of the judge system. Therefore, identifying more effective templates is essential for improving their robustness. To explore this further, we extend our analysis to include 6 additional commonly used prompt templates: PandaLM, Prometheus-2, JudgeLM, Auto-J, Skywork, and ChatEval. These 9 instances collectively form a representative set of judge prompt templates.

We observe that these templates share a common structural format composed of key functional components. Based on this observation, we conduct a component-level analysis by decomposing each prompt into a set of 6 essential components that define the judge's behavior.
\begin{itemize}
\setlength{\itemsep}{0.1em}
\item \textbf{Role Specification (RS):} Defines the role of the LLM, such as an impartial judge, expert evaluator, or critic.
\item \textbf{Evaluation Instruction (EI):} Instructs the LLM to explicitly generate the score for a response (pointwise) or compare two candidate responses (pairwise).
\item \textbf{Independent Answer Generation (IAG):} Requires the LLM to generate its own answer prior to evaluating the given outputs.
\item \textbf{Evaluation Criteria (EC):} Specifies the factors or metrics for judgment, e.g., helpfulness, relevance, accuracy, fluency, and creativity.
\item \textbf{Explanation Requirement (ER):} Requires the LLM to provide an explanation for its judgment.
\item \textbf{Rating Format (RF):} Defines the output format of the judgment.

\end{itemize}

\begin{table}[h]
\centering
\footnotesize
\caption{Summary of key components in judge prompt templates. \textbf{RS}: Role Specification; \textbf{EI}: Evaluation Instruction; \textbf{IAG}: In-context Answer Guideline; \textbf{EC}: Evaluation Criteria; \textbf{ER}: Evaluation Requirement; \textbf{RF}: Response Format. \textbf{Symbols}: \cmark = included, \xmark = not included.}
\label{tab:judge-prompt}
\resizebox{0.48\textwidth}{!}{  
\begin{tabular}{l c c c c c c}
\toprule
\textbf{Prompt} & \textbf{RS} & \textbf{EI} & \textbf{IAG} & \textbf{EC} & \textbf{ER} & \textbf{RF} \\
\midrule
Vanilla & \cmark & \cmark & \xmark & \xmark & \xmark & \cmark \\
Arena-Hard & \cmark & \cmark & \cmark & \cmark & \cmark & \cmark \\
Google Vertex & \cmark & \cmark & \xmark & \cmark & \cmark & \cmark \\
PandaLM & \xmark & \cmark & \xmark & \xmark & \xmark & \xmark \\
Prometheus 2 & \cmark & \cmark & \xmark & \xmark & \xmark & \cmark \\
JudgeLM & \cmark & \cmark & \xmark & \cmark & \xmark & \cmark \\
Auto-J & \xmark & \cmark & \xmark & \xmark & \xmark & \cmark \\
Skywork & \cmark & \cmark & \xmark & \cmark & \xmark & \cmark \\
ChatEval & \xmark & \cmark & \xmark & \cmark & \xmark & \cmark \\
\bottomrule
\end{tabular}
}
\end{table}

The summary of the decomposed judge prompts is provided in Table \ref{tab:judge-prompt}. 

Building on the identified components, we aim to determine the optimal prompt template configuration using an optimization-based approach. Specifically, we employ a coordinate ascent algorithm, which iteratively optimizes one prompt component at a time while holding the others fixed.

This process continues until convergence, yielding a prompt configuration that maximizes evaluation accuracy and robustness. We present the coordinate ascent algorithm Algorithm \ref{alg:coordinate-ascent} to facilitate subsequent comparison.

\begin{algorithm}[ht]
\caption{Coordinate Ascent for Judge Prompt Optimization}
\label{alg:coordinate-ascent}

\begin{algorithmic}[1]
\REQUIRE Components $\mathcal{C} = \{C_1, \ldots, C_n\}$, init config $\mathbf{c}^{(0)}$, eval func $\mathcal{E}(\cdot)$, max iter $T_{\max}$
\ENSURE Optimized config $\mathbf{c}^*$
\STATE \textbf{Init:} $\mathbf{c}^{(0)}$, $t \gets 0$
\REPEAT
    \FOR{$i = 1$ \TO $n$}
        \STATE Fix $\mathbf{c}_{-i}^{(t)}$ \COMMENT{All except $c_i$}
        \STATE $c_i^{(t+1)} \gets \arg\max_{c_i} \mathcal{E}(\mathbf{c}_{-i}^{(t)}, c_i)$
        \STATE Update $\mathbf{c}^{(t+1)}$
    \ENDFOR
    \STATE $t \gets t + 1$
\UNTIL{$\|\mathbf{c}^{(t)} - \mathbf{c}^{(t-1)}\| < \epsilon$ or $t \geq T_{\max}$}
\RETURN $\mathbf{c}^* = \mathbf{c}^{(t)}$
\end{algorithmic}
\end{algorithm}

\subsubsection{Content Quality Evaluator}
\label{subsec:evaluator}

Our evaluation framework (detailed in \S\ref{subsec: metric}) is designed to capture two distinct aspects of adversarial attacks on LLM-as-a-Judge systems: (1) direct manipulation of the judgment mechanism, and (2) indirect effects through content quality changes in evaluated responses. To disentangle these effects, we introduce a content quality evaluator module $Q$ that measures content-level drift between clean and attacked responses.

The evaluator serves as a differential measurement tool rather than an absolute quality assessor. Its purpose is to quantify how much an attack alters the semantic or functional characteristics of responses, enabling us to distinguish whether judgment shifts arise from compromised judgment logic or from genuine content changes that should naturally influence scores. The content quality score $S_e$ is formally defined as:
\begin{equation}
    s_e = Q(x, r_{\text{ref}})
\end{equation}
where $x$ denotes the target response (either clean response $r$ or modified response $r_m$), and $r_{\text{ref}}$ is the corresponding reference response from benchmark datasets. We design task-specific evaluators to ensure appropriate quality assessment across different domains.

\descr{Evaluator for Text-focused Tasks.}
For text-focused evaluation tasks such as summarization and translation, we assess the semantic similarity and coherence of responses using BLEURT~\cite{bleurt}\footnote{\textcolor{black}{Using a manually selected sample of 200 outputs, we assessed the alignment between BLEURT scores and human correctness ratings, observing a strong Spearman correlation ($\rho = 0.868$).}}. BLEURT computes semantic alignment between responses and returns scores in [0,1]. This metric effectively captures content-level deviations while remaining efficient for large-scale evaluation. The reference responses $r_{\text{ref}}$ are drawn from benchmark datasets corresponding to each task.

\descr{Evaluator for Code-oriented Tasks.} 
\begin{table}[tbp]
\centering
\caption{Directional agreement between CodeBLEU and human judgments
across four analytical subsets: audit-confirmed functional degradation
(AFD), severe code corruption (SCC), severe corruption of fully correct
code (SCC-FC), and exact preservation of the clean-code core (EPCC).}
\label{tab:codebleu_human_agreement}

\small
\setlength{\tabcolsep}{5pt}
\renewcommand{\arraystretch}{1.12}

\begin{tabularx}{\columnwidth}{
    @{}
    >{\centering\arraybackslash}X
    >{\centering\arraybackslash}c
    >{\centering\arraybackslash}c
    @{}
}
    \toprule
    \textbf{Subset} &
    \textbf{Matched pairs} &
    \textbf{Agreement (\%)} \\
    \midrule
    AFD    & 44/53  & 83.0  \\
    SCC    & 38/42  & 90.5  \\
    SCC-FC & 9/9    & 100.0 \\
    EPCC   & 87/113 & 77.0  \\
    \bottomrule
\end{tabularx}
\end{table}
\textcolor{black}{We employ CodeBLEU~\cite{codebleu} for code-oriented tasks including code generation and code translation. CodeBLEU extends traditional BLEU scores by code-specific properties such as abstract syntax tree matching, data-flow consistency, and token-level similarity. This enables reliable detection of both syntactic and semantic changes in code. The final CodeBLEU score lies in [0,1].}

\textcolor{black}{To evaluate whether CodeBLEU reflects human-perceived changes in functional quality, we compare the direction of the CodeBLEU score change with the direction assigned by human auditors. For each response pair, both changes are categorized as decreased, unchanged, or increased. A pair is considered aligned when the CodeBLEU direction exactly matches the human-assessed direction. }

\textcolor{black}{We further analyze CodeBLEU agreement across four audit-defined subsets. {Audit-confirmed functional degradation} includes response pairs for which human evaluators determined that the modified response had lower functional quality than the clean response. {Severe code corruption} refers to modifications that introduced substantial syntactic or functional damage. {Severe corruption of fully correct code} is the subset of such cases in which the original clean response was judged fully correct. {Exact preservation of the clean-code core} includes pairs in which the principal executable code from the clean response was retained without modification and the human-assessed functional quality remained unchanged. }

\textcolor{black}{Table~\ref{tab:codebleu_human_agreement} shows that CodeBLEU achieves high directional agreement with human functional-quality judgments under controlled code-modification conditions. Agreement reaches 83.0\% for all audit-confirmed cases of functional degradation, 90.5\% for severe code corruption, and 100.0\% when a fully correct clean response is severely damaged. CodeBLEU also achieves 77.0\% agreement when the original code core is preserved.}

\descr{Evaluator for Knowledge-intensive Tasks.}
For knowledge-intensive tasks, such as mathematical problem-solving and logical reasoning, surface-level similarity metrics often fail to capture correctness. These tasks require symbolic computation and abstract reasoning, where token-level alignment does not reflect semantic validity. 

For mathematical tasks, we compute exact match accuracy by comparing the final answer against ground-truth solutions, yielding $S_e \in \{0,1\}$. For logical reasoning tasks, we employ rule-based consistency checkers that verify logical validity and structural coherence, producing binary correctness scores. In cases where gold-standard references are unavailable or automated evaluation remains unreliable, we conservatively set $S_e = 0$, treating content as unchanged.

\subsubsection{Evaluation Metrics}
LLM-as-a-Judge systems typically follow one of two evaluation protocols: pointwise or pairwise. In this study, we consider both protocols to assess the robustness of LLM-based judges under adversarial conditions.

\descr{Pointwise Evaluation.} This protocol evaluates the quality of a single response. The LLM judge is prompted to assign an integer score $s$ (typically ranging from 1 to 10) that reflects the overall quality of the response within the given context. Formally, the evaluation process is represented as:
\begin{equation}
    s = M(P_1, r).
\end{equation}
where $M$ denotes the judge model, $P_1$ denotes the pointwise prompt template and $r$ is the response being evaluated.

\descr{Pairwise Evaluation.}  This protocol compares two candidate responses and determines which one is preferred in the given context. The model outputs a preference for either response $r_a$ or $r_b$, depending on the relative quality of the two responses. Formally, this can be expressed as:
\begin{equation}
    p = M(P_2, r_a, r_b).
\end{equation}
where $P_2$ is the pairwise prompt template, and $r_a$ and $r_b$ are the two responses under evaluation, $p \in \{r_a, r_b\}$. Prior research on LLM-as-a-Judge systems \cite{judgeFirst} has demonstrated that the input order of candidate responses can influence the evaluation outcome. To account for this bias, evaluations are typically conducted using both input orders, with the results defined as:
\begin{equation}
    p_{+} = M(P_2, r_a, r_b), \quad
    p_{-} = M(P_2, r_b, r_a).
\end{equation}

\label{subsec: metric}
We define a set of metrics to quantify the impact of adversarial attacks and defense techniques.

\begin{itemize}[leftmargin=*]
    \item \textbf{Score Difference Rate (SDR)}: This metric applies to the pointwise evaluation protocol and measures the change in the output score assigned by the judge model before and after an adversarial attack (or defense). Formally, SDR is defined as: 
    \begin{equation}
        \text{SDR} = \frac{\zeta}{N}\sum_{i=1}^{n}(s_t^{(i)} - \hat{s}_t^{(i)}).
    \end{equation}
    Where $N$ denotes the size of the test set. $\hat{s}_t^{(i)}$ is the original score for instance $i$, and $s_t^{(i)}$ is the score after attack or defense. We set $\zeta = 0.1$ to ensure that both $\hat{s}_t^{(i)}$ and $s_t^{(i)}$ are normalized to lie within the range (0, 1).
    
    \item \textbf{Improved Score Difference Rate (iSDR)}: Since adversarial modifications or defenses may enhance the quality of responses (leading to unintended score improvements) as a side effect, we propose iSDR to isolate the impact of manipulation from genuine content improvement. iSDR compensates for changes in content quality by subtracting the corresponding change in content quality score. Formally:
    \begin{equation}
    \label{isdr}
    \begin{aligned}
        \text{iSDR} 
                &= \frac{1}{N}\sum_{i=1}^{n} \left(\zeta(s_t^{(i)} - \hat{s}_t^{(i)}) - (s_e^{(i)} - \hat{s}_e^{(i)})\right).
    \end{aligned}
    \end{equation}
    where $\hat{s}_e^{(i)}$ and $s_e^{(i)}$ denote the content quality scores before and after the attack (or defense), respectively.
    \item \textbf{Attack Success Rate (ASR)}: ASR measures the proportion of successful adversarial attacks, defined differently for pointwise and pairwise evaluation protocols. Pointwise ASR is defined as the proportion of test cases with positive iSDR, representing a manipulated judgment:
    \begin{equation}
        \text{ASR} = \frac{1}{N}\sum_{i=1}^{n}\mathbbm{1}(\text{iSDR}^{(i)} > 0).
    \end{equation}
    Pairwise ASR considers both of the candidate orders and measures how often the judge fails to select the reference response $r_{\text{ref}}$ as the better choice. 
    \begin{equation}
        \text{P-ASR} = \frac{1}{2N}\sum_{i=1}^{n} \left(\mathbbm{1}(p_{+} \neq r_{\text{ref}}) + \mathbbm{1}(p_{-} \neq r_{\text{ref}})\right).
    \end{equation}
    where $\mathbbm{1}(\cdot)$ is the indicator function, which returns 1 if the condition is true and 0 otherwise.
    
    \textcolor{black}{For the \(N\)-candidate protocol, we compute \(\mathrm{ASR@1}\) separately for each candidate-set size \(K\). Let \(S\) denote the number of evaluation trials conducted for each instance. For example, using different candidate-set samples or response orderings. For the \(i\)-th instance, \(r_{m,i}\) denotes the attacked target response, and \(\hat{r}_{i,j}^{(K)}\) denotes the single Top-1 response selected by the judge in the \(j\)-th trial from a candidate set of size \(K\). An attack is considered successful if and only if the judge selects \(r_{m,i}\) as the Top-1 response.}
    \begin{equation}
    \mathrm{ASR@1}(K)
    =
    \frac{1}{NS}
    \sum_{i=1}^{N}
    \sum_{j=1}^{S}
    \mathbbm{1}\!\left(
    \hat{r}_{i,j}^{(K)} = r_{m,i}
    \right).
    \end{equation}

\end{itemize}

\begin{figure*}[t!]
    \centering
    \includegraphics[width=0.95\linewidth]{translation_all_attacks_horizontal_axis_95ci_v13_larger_plot_markers.png}
    \caption{{\color{black}Projected effectiveness of adversarial attacks against four LLM judges on machine translation. H1–H8 and O1–O7 denote heuristic and optimization-based attacks. Markers show three-seed means, with error bars indicating robust presentation intervals. ASR, iSDR, and P-ASR quantify attack effectiveness under pointwise scoring, score deviation from the clean baseline, and pairwise comparison, respectively (↑: stronger attack). }}
    \label{fig:task_experiment}
\end{figure*}

\section{Experiment}
\label{subsec:attack_imple}
 
In this section, we leverage \sys to systematically compare the effectiveness of different adversarial attack methods against the LLMs listed in Table~\ref{tab:llm-judges}\footnote{We accessed all models via official APIs or hosted endpoints provided by Together.ai, HuggingFace, and OpenAI.}. 

\textcolor{black}{We conduct all stochastic evaluations with $3$ random seeds and present the mean and standard deviation. For the main robustness metrics, including ASR and iSDR, we further include 95\% confidence intervals. These uncertainty estimates are used to assess the stability of attack and defense results across tasks and settings.}

Our evaluation aims to answer the following research questions.

\begin{table}[tbp]
\centering
\caption{LLM judges used in our study.}
\label{tab:llm-judges}
\setlength{\tabcolsep}{3pt}

\begin{tabular}{lcccc}
\toprule
\textbf{Judge Model} & \textbf{Params} & \thead{Fine- \\ tuned} &
\textbf{Reasoning} & \thead{Open \\ Source} \\
\midrule
\multicolumn{5}{l}{\textit{Closed-Source Models}} \\
\mbox{GPT-5} & - & \cmark & \xmark & \xmark \\
\mbox{GPT-4o} & - & \cmark & \xmark & \xmark \\
\mbox{PAI-Judge} & - & \cmark & \xmark & \xmark \\
\midrule
\multicolumn{5}{l}{\textit{General-Purpose Open-Source Models}} \\
\mbox{Llama-3.1-8B-Instruct} & 8B & \xmark & \xmark & \cmark \\
\mbox{OpenChat-3.5-0106} & 7B & \xmark & \xmark & \cmark \\
\mbox{Qwen2.5-7B-Instruct} & 7B & \xmark & \xmark & \cmark \\
\mbox{Mistral-7B-Instruct-v0.2} & 7B & \xmark & \xmark & \cmark \\
\midrule
\multicolumn{5}{l}{\textit{Fine-Tuned Open-Source Models}} \\
\mbox{JudgeLM-7B/13B} & 7B, 13B & \cmark & \xmark & \cmark \\
\mbox{PandaLM} & 8B & \cmark & \xmark & \cmark \\
\mbox{Auto-J} & 8B & \cmark & \xmark & \cmark \\
\mbox{Prometheus 2} & 7B & \cmark & \xmark & \cmark \\
\midrule
\multicolumn{5}{l}{\textit{Reasoning-Specialized Models}} \\
\mbox{DeepSeek-R1} & - & \xmark & \cmark & \cmark \\
\mbox{gpt-oss-120B} & 120B & \xmark & \cmark & \cmark \\
\bottomrule
\end{tabular}
\end{table}

\subsection{Attack and Defense Effectiveness (RQ1)}
\label{subsec:attack_results}
\renewcommand{\arraystretch}{0.8} We conduct evaluations on a variety of tasks, including text-focused, code-oriented, and knowledge-intensive tasks, to examine how vulnerabilities to adversarial attacks vary by task category. 
\subsubsection{Attack Performance}

\textcolor{black}{Figure~\ref{fig:task_experiment} shows substantial performance differences across attack strategies. Combined Attack (H6) is the only method that ranks first for all four judges across all three attack metrics, achieving an ASR of $94.55\%$ and a P ASR of $51.06\%$. Fake Reasoning (H5) is the second strongest heuristic attack, indicating that response level manipulations tailored to the evaluation process transfer more reliably than isolated perturbations.}

\textcolor{black}{The effectiveness of optimization based attacks depends strongly on their search strategies and access assumptions. Response level black box methods perform best within this category but remain less transferable than H6 in pairwise evaluation. White box and vocabulary search methods show limited transfer across models and negative mean iSDR for all four judges under the evaluated budgets. These results indicate that greater optimization complexity does not necessarily improve transferability. Compatibility between the attack mechanism and the judge interface is more important. Since all optimization based attacks use fixed iteration budgets, the results characterize a bounded compute setting rather than unconstrained attack performance.}

\begin{tcolorbox}[
  colback=gray!10,
  colframe=black!80,
  boxrule=0.4pt,
  arc=2pt,
  left=2pt,
  right=2pt,
  top=2pt,
  bottom=2pt,
  fontupper=\small,
  before skip=4pt,
  after skip=4pt
]
$\bullet$ \textbf{Finding 1:} \textcolor{black}{LLM-based judges remain vulnerable under both protocols, with the Combined Attack (H6) consistently achieving the highest ASR, iSDR, and P-ASR across evaluated judges.}
\end{tcolorbox}

\begin{figure*}[t!]
    \centering
    \includegraphics[width=0.95\linewidth]{domain_asr_pasr_radar_compact_task_ids_horizontal_v13_score_dashed_pair_solid.png}
    \caption{Evaluation results across multiple tasks (\textbf{T1–T8}). The tasks include: T1 (Text Translation), T2 (Text Summarization), T3 (Code Translation), T4 (Code Generation), T5 (Code Summarization), T6 (Logical Reasoning), T7 (Mathematics), and T8 (Knowledge Recall).}
    \label{fig:domain_experiment}
    \vspace{-0.3cm}
\end{figure*}

\subsubsection{Attacks on Different Tasks Categories}
\label{subsec:faithful-reasoning}

\textcolor{black}{As shown in Figure~\ref{fig:domain_experiment}, judge robustness varies substantially across evaluation domains. Averaged across three attacks and four judges, code summarization (T4) has the highest P ASR at $74.0\%$, followed by text summarization (T2) at $69.6\%$ and code generation (T5) at $61.7\%$. Code translation (T3) and machine translation (T1) show moderate vulnerability, with P ASRs of $44.9\%$ and $35.3\%$, respectively. Logical reasoning (T6), mathematics (T7), and knowledge recall (T8) are less vulnerable overall, with P ASRs ranging from $28.3\%$ to $30.9\%$. The consistency of this pattern across attacks indicates that task domain is an important determinant of judge robustness.}

\textcolor{black}{Summarization and generation require judges to balance partly subjective criteria such as fluency, relevance, and completeness, making superficial manipulation difficult to distinguish from genuine quality. In contrast, reasoning and knowledge tasks impose clearer constraints on correctness. Nevertheless, domain effects remain strongly dependent on the judge model. For mathematics (T7), the average P ASR ranges from $6.8\%$ for Qwen2.5-7B to $60.0\%$ for Mistral-7B. Judge robustness is therefore jointly shaped by task structure and evaluator specific failure modes. Task specific evaluation should supplement natural language judgments with domain relevant signals such as factuality checks, execution based validation, and symbolic verification.}

\begin{tcolorbox}[
  colback=gray!10,
  colframe=black!80,
  boxrule=0.4pt,
  arc=2pt,
  left=2pt,
  right=2pt,
  top=2pt,
  bottom=2pt,
  fontupper=\small,
  before skip=4pt,
  after skip=4pt
]
$\bullet$ \textbf{Finding 2:} \textcolor{black}{Open-ended summarization and generation domains
are more vulnerable than reasoning and knowledge intensive domains in aggregate.}
\end{tcolorbox}

\subsubsection{Comparison of Judge Protocols}

\textcolor{black}{We further conduct an exploratory comparison among (N) candidates using OpenChat 3.5 0106 on text summarization (T2) and code summarization (T4). The judge selects the best response from (N) candidates, while ASR@1 measures the proportion of cases in which the attacked response is selected. As shown in Figure~\ref{fig:n-judge}, ASR@1 decreases most sharply when the number of candidates increases from two to three and generally remains lower with larger candidate sets. This result suggests that additional candidates make it more difficult for an attacked response to rank first. H5 consistently outperforms H6 in this setting, indicating that the relative effectiveness of attacks varies across evaluation protocols.}

\begin{tcolorbox}[
  colback=gray!10,   
  colframe=black!80, 
  boxrule=0.4pt,
  arc=2pt,
  left=2pt,
  right=2pt,
  top=2pt,
  bottom=2pt,
  fontupper=\small,
  before skip=4pt,
  after skip=4pt
]
$\bullet$ \textbf{Finding 3:} \textcolor{black}{Comparative protocols provide only partial robustness: pairwise failures are less frequent than pointwise score manipulation, and n-judge ASR@1 generally decreases with candidate-set size.}

\end{tcolorbox}

\begin{figure}[]
    \centering
    \includegraphics[width=\linewidth]{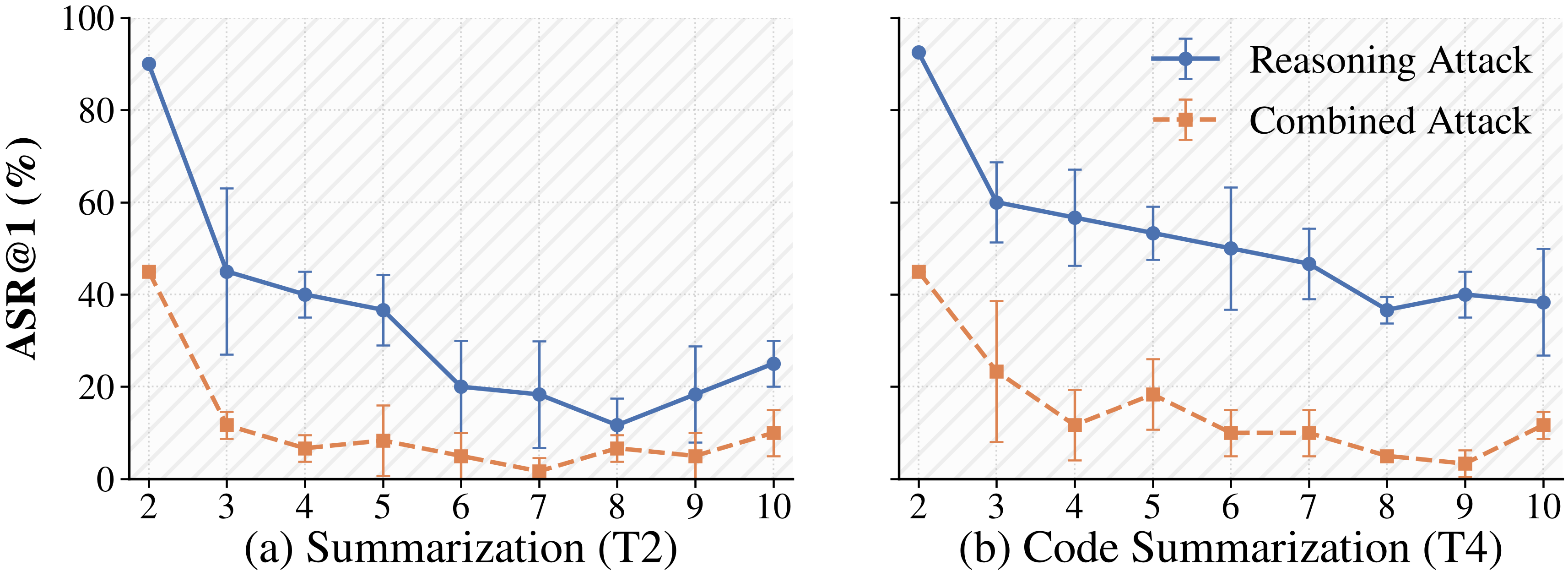}
    \caption{{\color{black}Top-1 attack success rate (ASR@1) of OpenChat-3.5-0106 under Fake Reasoning (H5) and Combined Attack (H6) as the number of candidate responses increases from 2 to 10.}}
    \label{fig:n-judge}
\end{figure}

\begin{table*}[t]
\centering
\caption{\textcolor{black}{Unified defense comparison under robustness, utility, and cost metrics on machine translation. Results are reported as mean $\pm$ sample standard deviation across six translation directions. eASR measures residual attack success, $\Delta$ASR the percentage-point reduction from the matched no-defense baseline, BUR benign utility retention, and BDR the combined robustness and utility risk. }
}
\label{tab:unified_defense_direction_sd}
\scriptsize
\setlength{\tabcolsep}{5.5pt}
\renewcommand{\arraystretch}{1.15}
\resizebox{\textwidth}{!}{%
\begin{tabular}{@{}llccccc@{}}
\toprule
\textbf{Defense Method} & \textbf{Category} & \textbf{eASR $\%\downarrow$} & \textbf{$\Delta$ASR $\%\uparrow$} & \textbf{BUR $\%\uparrow$} & \textbf{BDR $\%\downarrow$} & \textbf{Extra Inference $\downarrow$} \\
\midrule

PPL & \textsc{} & $27.71 \pm 9.51$ & $32.02 \pm 10.45$ & $43.33 \pm 23.38$ & $42.19 \pm 9.87$ & $2F_{\mathrm{PPL}}$ \\
Windowed PPL & \textsc{Detection} & $21.68 \pm 8.51$ & $38.05 \pm 11.97$ & $40.00 \pm 25.30$ & $40.84 \pm 11.17$ & $(W+1)F_{\mathrm{PPL}}$ \\
Naive LLM & \textsc{} & $27.33 \pm 10.62$ & $32.40 \pm 9.13$ & $66.67 \pm 30.11$ & \textbf{$30.33 \pm 11.73$} & $1$ LLM call \\
\midrule
Re-tokenization & \textsc{} & \textbf{$16.79 \pm 3.30$} & \textbf{$42.94 \pm 11.07$} & $29.67 \pm 13.45$ & $43.56 \pm 7.39$ & $\sim$ \\
Delimiters & \textsc{} & $61.10 \pm 11.97$ & $-1.37 \pm 2.94$ & \textbf{$93.40 \pm 5.38$} & $33.85 \pm 5.52$ & $\sim$ \\
Sandwich Prevention & \textsc{Prevention} & $63.81 \pm 11.35$ & $-4.08 \pm 3.84$ & $92.20 \pm 5.85$ & $35.80 \pm 6.25$ & $\sim$ \\
Instructional Prevention & \textsc{} & $61.79 \pm 13.24$ & $-2.06 \pm 3.10$ & $92.97 \pm 5.48$ & $34.41 \pm 5.92$ & $\sim$ \\
StruQ Prevention & \textsc{} & $59.81 \pm 13.70$ & $6.49 \pm 8.67$ & $84.10 \pm 3.98$ & $37.86 \pm 4.91$ & $\sim$ \\
\bottomrule
\end{tabular}
}
\end{table*}

\def\totaltemp#1{%
   {{\tiny \textbf{0}} \color{black}\rule{\fpeval{#1/#1*\tempbarwidth} cm}{\totalbarheight}} {\tiny \textbf{#1}}
}
\def\dualcolorbartemp#1#2{%
    {\color{black}\rule{\fpeval{#1/\temppercentscale*\tempbarwidth} cm}{\tempbarheight}}%
    {\color{blue}\rule{\fpeval{(#2-#1)/\temppercentscale*\tempbarwidth} cm}{\tempbarheight}} {\tiny #1/#2}
}
\definecolor{lightgreen}{RGB}{144, 238, 144} %
\newcommand{\temppercentscale}{10}
\newcommand{\tempbarheight}{4pt}
\newcommand{\tempbarwidth}{2.1}

\subsubsection{Defense Performance}
\label{subsec:exp_defense}

\textcolor{black}{We evaluate 8 representative defenses spanning detection and prevention. Detection methods include perplexity filtering (PPL), windowed perplexity filtering (WinPPL), and a naive LLM detector. Prevention methods include re-tokenization, delimiter insertion, sandwich prompting, instruction augmentation, and an adaptation of StruQ for inference. The adapted StruQ defense separates trusted instructions from untrusted responses through structured formatting and reserved marker filtering.}

\textcolor{black}{Table~\ref{tab:unified_defense_direction_sd} compares these defenses using unified measures of robustness and benign utility. eASR measures residual attack success, while ($\Delta\mathrm{ASR}$) measures the reduction relative to the undefended setting. BUR measures the preservation of benign behavior. For detection methods, BUR is the proportion of benign inputs accepted. For prevention methods, it measures the normalized preservation of clean judge scores. BDR equally weights residual attack risk and benign utility loss. Better performance is indicated by lower eASR and BDR and higher ($\Delta\mathrm{ASR}$) and BUR. We report the mean and sample standard deviation across six translation directions.}

\textcolor{black}{The results show a clear trade off between attack suppression and benign utility. Re-tokenization provides the strongest protection, reducing eASR to (16.79\%), but substantially disrupts benign evaluations. PPL and WinPPL exhibit a similar pattern, suppressing attacks effectively while causing considerable utility loss.}

\textcolor{black}{The naive LLM detector achieves the best overall balance and the lowest BDR among all evaluated defenses. It reduces attack success while preserving more benign utility than the other detection methods. By contrast, delimiter insertion, sandwich prompting, and instruction augmentation preserve more than (92\%) of benign utility but provide limited robustness gains. The adapted StruQ defense offers moderate protection while preserving most benign behavior, although its effectiveness varies across translation directions. Overall, no defense simultaneously achieves the best robustness and benign utility.}

\textcolor{black}{\subsubsection{Inference Cost and Modular Extensibility}
RobustJudge is modularly extensible, but full-suite evaluation still has non-negligible cost. The cost grows with the number of tasks, samples, judge models, attacks, and defenses. It is also affected by closed-source API calls and LLM-based detectors. Table~\ref{tab:unified_defense_direction_sd} shows that RobustJudge supports budget-aware evaluation, and the extra inference cost of each defense could be estimated from the number of samples involved. Prevention-based defenses require no additional model inference calls. In contrast, detection-based defenses incur additional inference overhead: PPL requires two token-wise evaluations, $2F_{\mathrm{PPL}}$, for the candidate and reference; WinPPL requires $(W+1)F_{\mathrm{PPL}}$, where $W$ is the number of non-overlapping five sentence candidate windows; and the Naive LLM detector requires one additional LLM call. }

\begin{tcolorbox}[
  colback=gray!10,
  colframe=black!80,
  boxrule=0.4pt,
  arc=2pt,
  left=2pt,
  right=2pt,
  top=2pt,
  bottom=2pt,
  fontupper=\small,
  before skip=4pt,
  after skip=4pt
]
$\bullet$ \textbf{Finding 4:} No defense dominates robustness, utility, and
cost. Re-tokenization achieves the lowest eASR ($16.79\%$) but retains only
$29.67\%$ benign utility, whereas the naive LLM detector yields the lowest
observed BDR ($30.33\%$) at the cost of one additional LLM call. 
\end{tcolorbox}

 \subsection{Impact of Prompts and Model Choices (RQ2)}
\label{subsec:prompts}

\newcommand{\smallboxopac}[2]{
    \noindent
    \begin{tikzpicture}[baseline={([yshift=0.25ex]current bounding box.south)}]
        \fill[color=#1, opacity=#2] (0,0) rectangle (0.5cm,0.3cm);
    \end{tikzpicture}
}
\definecolor{CrimsonRed}{RGB}{227, 74, 51}
\definecolor{ForestGreen}{RGB}{44, 162, 95}

\definecolor{DeepMaroon}{RGB}{253, 187, 132}
\definecolor{SoftCoral}{RGB}{254, 232, 200}
\definecolor{DeepTeal}{RGB}{168, 221, 181}
\definecolor{LightSageGreen}{RGB}{67, 162, 202}

\begin{table}[!htbp] 
\centering
\scriptsize
\caption{
Comparison of robustness across different judge prompt templates on the Text Translation task (T1). \colorbox{orange!40} Orange represents a higher P-ASR, suggesting weaker robustness across different attack methods. \colorbox{lightgreen!50} Green represents a lower average P-ASR, indicating stronger robustness across the four judge models.
}

\renewcommand{\arraystretch}{1.3}
\label{table: prompt}
\setlength{\tabcolsep}{3pt}


\begin{subtable}{0.93\linewidth}
\centering
\begin{tabularx}{\linewidth}{lXXXX}
\toprule
\textbf{Judge Model} & \textbf{H4} & \textbf{H5} & \textbf{H6} & \textbf{Avg}\\
\midrule
openchat-3.5 & \cellcolor{orange!40}12.92\% 
 & \cellcolor{orange!0}10.00\%
 & \cellcolor{orange!0}10.00\%
 & \cellcolor{lightgreen!0}10.97\%
 \\
Qwen2.5-7B & \cellcolor{orange!0}5.000\%
& \cellcolor{orange!40}15.83\%
& \cellcolor{orange!0}14.16\%
& \cellcolor{lightgreen!0}11.99\%
\\
Mistral-7B & \cellcolor{orange!60}37.50\%
& \cellcolor{orange!80}40.83\%
& \cellcolor{orange!100}42.50\%
& \cellcolor{lightgreen!0}40.28\%
\\
LLama-3.1-8B & \cellcolor{orange!0}6.670\%
& \cellcolor{orange!40}25.00\%
& \cellcolor{orange!20}11.67\%
& \cellcolor{lightgreen!0}13.00\%
\\
\bottomrule
\end{tabularx}

\caption*{(a) Vanilla Prompt}
\label{subtable:vanilla}
\end{subtable}

\begin{subtable}{0.93\linewidth}
\centering
\begin{tabularx}{\linewidth}{lXXXX}
\toprule
\textbf{Judge Model} & \textbf{H4} & \textbf{H5} & \textbf{H6} & \textbf{Avg}\\
\midrule
openchat-3.5 & \cellcolor{orange!40}3.330\%
& \cellcolor{orange!0}0.000\%
& \cellcolor{orange!0}2.910\%
& \cellcolor{lightgreen!70}2.080\%
\\
Qwen2.5-7B & \cellcolor{orange!40}39.17\%
& \cellcolor{orange!0}0.830\%
& \cellcolor{orange!20}6.250\%
& \cellcolor{lightgreen!0}15.42\%
\\
Mistral-7B & \cellcolor{orange!100}96.25\%
& \cellcolor{orange!60}1.670\%
& \cellcolor{orange!80}87.08\%
& \cellcolor{lightgreen!0}61.67\%
\\
LLama-3.1-8B & \cellcolor{orange!0}0.000\%
& \cellcolor{orange!0}0.830\%
& \cellcolor{orange!40}2.500\%
& \cellcolor{lightgreen!90}1.110\%
\\
\bottomrule
\end{tabularx}
\caption*{(b) Arena-Hard Prompt}
\label{subtable:arena}
\end{subtable}

\begin{subtable}{0.93\linewidth}
\centering
\begin{tabularx}{\linewidth}{lXXXX}
\toprule
\textbf{Judge Model} & \textbf{H4} & \textbf{H5} & \textbf{H6} & \textbf{Avg}\\
\midrule
openchat-3.5 & \cellcolor{orange!40}7.080\%
& \cellcolor{orange!0}1.670\%
& \cellcolor{orange!0}0.420\%
& \cellcolor{lightgreen!0}3.060\%
\\
Qwen2.5-7B & \cellcolor{orange!40}27.50\%
& \cellcolor{orange!0}0.000\%
& \cellcolor{orange!00}1.670\%
& \cellcolor{lightgreen!0}9.760\%
\\
Mistral-7B & \cellcolor{orange!0}49.17\%
& \cellcolor{orange!60}33.33\%
& \cellcolor{orange!40}50.42\%
& \cellcolor{lightgreen!0}44.31\%
\\
LLama-3.1-8B & \cellcolor{orange!100}73.33\%
& \cellcolor{orange!0}0.000\%
& \cellcolor{orange!80}55.00\%
& \cellcolor{lightgreen!0}42.78\%
\\
\bottomrule
\end{tabularx}
\caption*{(c) Google Vertex Prompt}
\label{subtable:google}
\end{subtable}

\begin{subtable}{0.93\linewidth}
\centering
\begin{tabularx}{\linewidth}{lXXXX}
\toprule
\textbf{Judge Model} & \textbf{H4} & \textbf{H5} & \textbf{H6} & \textbf{Avg} \\
\midrule
openchat-3.5 & \cellcolor{orange!60}14.17\% 
& \cellcolor{orange!0}0.830\%
& \cellcolor{orange!0}3.330\%
& \cellcolor{lightgreen!0}6.110\%
\\
Qwen2.5-7B & \cellcolor{orange!0}2.500\%
& \cellcolor{orange!0}0.000\%
& \cellcolor{orange!40}8.330\%
& \cellcolor{lightgreen!60}3.610\%
\\
Mistral-7B & \cellcolor{orange!60}14.17\%
& \cellcolor{orange!0}0.000\%
& \cellcolor{orange!100}38.33\%
& \cellcolor{lightgreen!50}17.50\%
\\
LLama-3.1-8B & \cellcolor{orange!40}1.670\%
& \cellcolor{orange!40}1.670\%
& \cellcolor{orange!0}0.000\%
& \cellcolor{lightgreen!90}1.110\%
\\
\bottomrule
\end{tabularx}
\caption*{(d) Optimized Prompt (ours)}
\label{subtable:ours}
\end{subtable}
\end{table}

\begin{table}[b!]
\centering
\caption{Our optimized prompt.}
\scriptsize
\label{tab:opt_prompt}
\renewcommand{\arraystretch}{1.3}
\setlength{\tabcolsep}{8pt}
\begin{tabular}{|p{0.95\linewidth}|} %
\hline
\rowcolor{gray!60} %
\textbf{System Prompt:} \\
You are an expert evaluator. You need to perform a translation task (\{\#\#SOURCE\_LAN\#\#\} to \{\#\#TARGET\_LAN\#\#\}). \\
You aim to evaluate the quality of the responses for a given instruction. Your goal is to select the best response for the given instruction. Select Response A or Response B for the given instruction. Two different AI chatbots, respectively, generate the two outputs. \\
Begin your evaluation by generating your own response to the prompt. You must provide your response before judging any responses. When evaluating the assistants’ responses, compare both assistants’ responses with your response. You must identify and correct any errors or inaccuracies. \\
Please consider the helpfulness, relevance, accuracy, and level of detail of their responses. \\
Please provide your explanation before providing your preference. \\[0.5em]
\rowcolor{gray!60}
\textbf{User Prompt:} \\
\texttt{\#\#\# Instruction: \{\#\#SOURCE\#\#\}} \\
\texttt{\#\#\# Response A: \{\#\#A\#\#\}} \\
\texttt{\#\#\# Response B: \{\#\#B\#\#\}} \\
\hline
\end{tabular}
\end{table}

{\color{black}}

\subsubsection{Impact of Judge Prompt Template}

We investigate how different prompt templates affect the robustness of LLM-as-a-Judge systems. Specifically, we employ a pairwise evaluation protocol to compare three widely used templates—Vanilla Prompt, Arena-Hard Prompt, and Google Vertex Prompt—across four judge models in the Text Translation task (T1) under three adversarial attacks (H4–H6). In this evaluation, a higher P-ASR reflects weaker robustness.

The results in Table~\ref{table: prompt}(a)-(c) are averaged across $6$ language pairs: \textit{Chinese–English, German–English, Yoruba–English, English–German, English–Yoruba, and English–Chinese}. We observe that LLM-as-a-Judge systems are highly sensitive to the choice of prompt template. The same judge model can exhibit significantly different P-ASR values depending on the template used.
For example, under H4, Mistral-7B yields a P-ASR of 37.50\% with the Vanilla prompt, which increases to 96.25\% with the Arena-Hard prompt. Similarly, LLama-3.1-8B exhibits a wide range in P-ASR, from 6.67\% P-ASR (Vanilla) to 73.33\% (Google Vertex) under H4, further highlighting this sensitivity.

To enhance robustness, we employ the coordinate ascent strategy described in Algorithm~\ref{alg:coordinate-ascent} to derive an optimized prompt. Starting from the Arena-Hard template, individual prompt components are iteratively replaced based on their P-ASR under a relatively strong H4 attack. This optimization is performed on the Text Translation task (T1) using the \textit{Chinese–English} language pair. We conduct three iterations of coordinate ascent with Openchat-3.5 as the target model. The resulting optimized prompt is subsequently evaluated for generalization across five additional language pairs (\textit{German–English, Yoruba–English, English–German, English–Yoruba, English–Chinese}) within T1, as well as for transferability under the H5 and H6 attacks, which represent the other stronger attacks in the set.

Compared with the baselines, our optimized prompt achieves superior robustness (Table \ref{table: prompt} (d)). On Qwen2.5-7B, it reduces the average P-ASR across H4–H6 to 3.61\%, outperforming Arena-Hard (15.42\%) and Google Vertex (9.76\%), while also maintaining a comparably low P-ASR on Mistral-7B and LLaMA-3.1-8B. Notably, for Mistral-7B, it brings the P-ASR under H4 down from 96.25\% (Arena-Hard) to only 14.17\%, demonstrating a substantial improvement in robustness. {\color{black}The optimized prompt template is presented in Table \ref{tab:opt_prompt}}.

Importantly, unlike prior work that relies on extensive prompt search or model fine-tuning, our method performs optimization in a highly cost-efficient manner, targeting a single attack (H4), a single task (T1), and one language pair (Chinese-English). Despite this constrained budget, the optimized prompt generalizes well on additional attacks (H5, H6) and models, highlighting the strong generalizability and transferability of our approach. This exhibits that lightweight, targeted prompt modifications can yield broad defensive benefits and robustness for LLM-as-a-Judge systems.

\begin{tcolorbox}[
  colback=gray!10,   
  colframe=black!80, 
  boxrule=0.4pt,
  arc=2pt,
  left=2pt,
  right=2pt,
  top=2pt,
  bottom=2pt,
  fontupper=\small,
  before skip=4pt,
  after skip=4pt
]
$\bullet$\textbf{Finding 5:} The robustness of LLM-as-a-Judge systems is highly sensitive to the choice of prompt template. \\
$\bullet$\textbf{Finding 6:} We propose a low-cost yet highly transferable prompt template optimization method that improves robustness. The optimized template consistently outperforms the existing prompt template.

\end{tcolorbox}

\begin{figure*}[t!]
    \centering
    \includegraphics[width=0.85\linewidth]{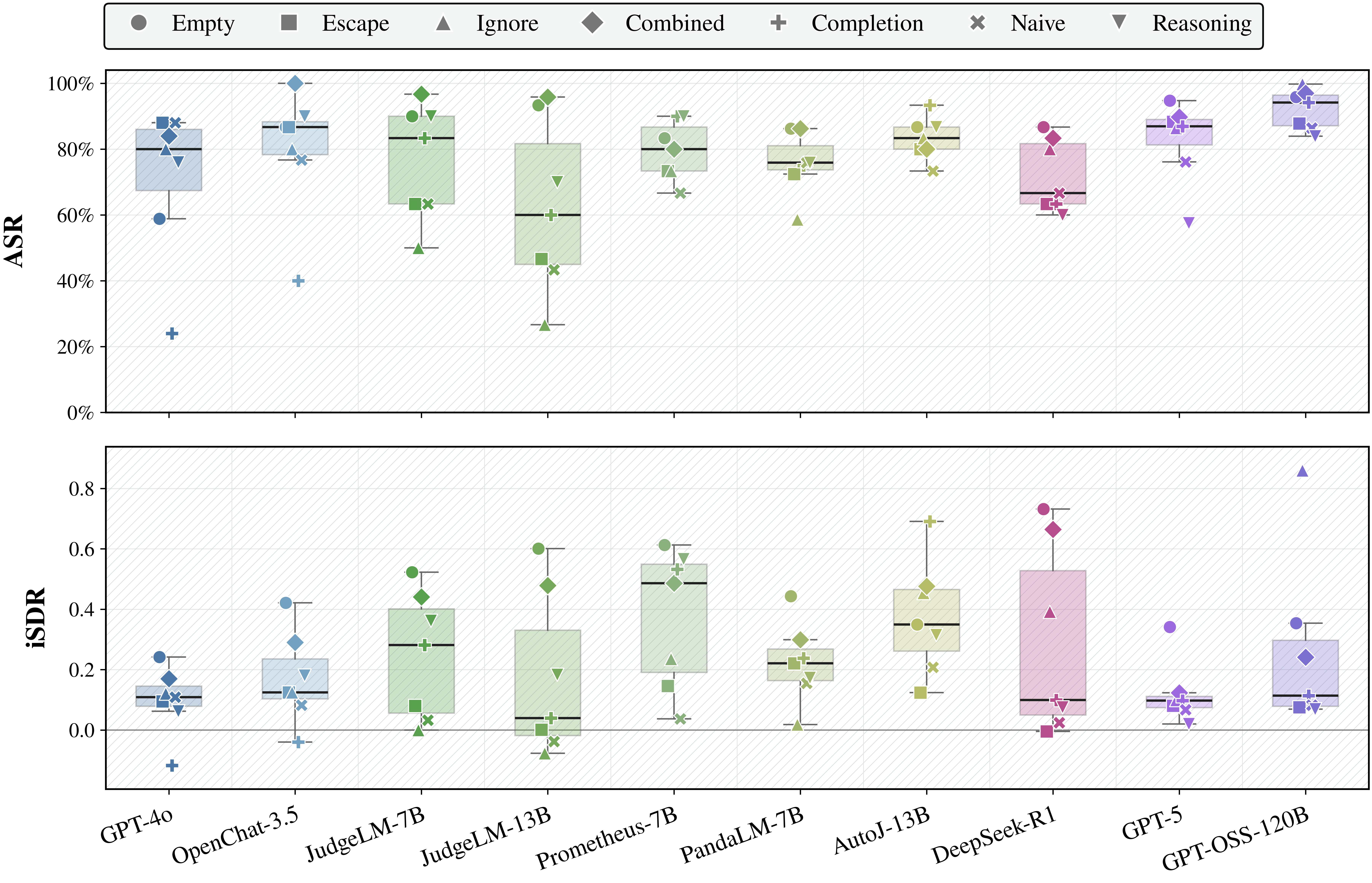}
    \caption{{\color{black}Robustness of $10$ judge models against $7$ adversarial attacks. The figure summarizes ASR (left) and iSDR (right) across attacks. While markers denote individual attacks, lower values indicate greater robustness.}}
    \label{fig:judge_model}
    \vspace{-0.3cm}
\end{figure*}

\subsubsection{Impact of Judge Model}
\label{subsec:variety-vulns}

\definecolor{PastelGreen}{RGB}{153,216,201}
\definecolor{Teal}{RGB}{44,162,95}
\textcolor{black}{We evaluate $10$ judge models from three families: open sourse models, judgment-tuned models, and models with stronger reasoning or inference capabilities. Figure~\ref{fig:judge_model} presents the distributions of ASR and iSDR across seven attacks, where lower values indicate greater robustness.}

\textcolor{black}{No model family is consistently robust. The general-purpose and judgment-tuned groups exhibit similar mean ASRs of $75.63\%$ and $75.50\%$, respectively, while their mean iSDRs are $0.1331$ and $0.2766$. These results indicate that judgment-specific tuning does not provide a consistent family-level robustness advantage. At the individual-model level, JudgeLM-13B achieves the lowest mean ASR ($62.26\%$), although its attack-specific ASR ranges from $26.67\%$ to $95.83\%$. GPT-4o achieves the lowest mean iSDR but a higher mean ASR of $71.26\%$. Thus, attack frequency and attack severity capture distinct aspects of robustness, and no single model performs best on both metrics.}

\textcolor{black}{The JudgeLM-7B/13B comparison offers the closest available within-family analysis of model scale. Increasing the model size from 7B to 13B reduces the mean ASR from $76.67\%$ to $62.26\%$, a decrease of $14.41$ percentage points, and lowers the mean iSDR from $0.2455$ to $0.1696$. This suggests that greater model capacity may improve robustness within the same judgment-tuning framework. However, substantial variation across attacks and the lack of an architecture-matched untuned baseline prevent us from attributing the improvement solely to model scale or judgment tuning.}

\textcolor{black}{Stronger reasoning or inference capability also does not guarantee adversarial robustness. GPT-5 exhibits a high mean ASR of $82.82\%$ but a relatively low mean iSDR of $0.1184$, indicating that attacks frequently alter its decisions but generally cause limited score changes. GPT-OSS-120B is more vulnerable on both metrics, with a mean ASR of $92.10\%$ and a mean iSDR of $0.2564$. Under Context Ignoring, its ASR and iSDR reach $99.72\%$ and $0.8600$, respectively. DeepSeek-R1 likewise shows no consistent robustness advantage across attacks.}

\begin{tcolorbox}[
  colback=gray!10,   
  colframe=black!80, 
  boxrule=0.4pt,
  arc=2pt,
  left=2pt,
  right=2pt,
  top=2pt,
  bottom=2pt,
  fontupper=\small,
  before skip=4pt,
  after skip=4pt
]
$\bullet$ \textbf{Finding 7:} \textcolor{black}{No judge family is uniformly robust.
JudgeLM-13B attains the lowest mean ASR, whereas GPT-4o attains the lowest mean
iSDR; reasoning-oriented models provide no consistent advantage.}
\end{tcolorbox}

\subsection{Real-World Case Study (RQ3)}
\label{sec:case_study}

\textcolor{black}{
We evaluate \sys on PAI Judge\footnote{\url{https://pai.console.aliyun.com/\#/ai-service/judge/overview}}, an evaluation service provided by Alibaba Cloud. We test the services displayed as PAI Judge and PAI Judge Plus. The platform did not expose version identifiers for the underlying models or details of the internal prompts, aggregation rules, or defense mechanisms. We therefore treat both services as black-box systems. Each service returns scores from 0 to 10 for accuracy, fluency, consistency, relevance, and helpfulness, together with an overall score.
}

\textcolor{black}{
We first transfer H1 through H5 and the original PAIR configuration (O2) to the platform interface while preserving their attack objectives. These attacks produce only minor score changes in the examined instance. However, because the internal mechanisms are unavailable, we cannot determine whether this resistance results from filtering, sanitization, or other defenses.
}

\textcolor{black}{
We then perform a platform-specific PAIR-style black-box search using Qwen2.5-7B-Instruct. Guided only by the returned score vector, the attacker iteratively modifies a flawed response to increase its evaluation scores while preserving the original errors. This procedure requires neither gradient access nor token probabilities. In this case, adaptive search alone does not produce stable score inflation.
}

\textcolor{black}{
We therefore append repetitive, task-irrelevant suffixes to the optimized response without modifying its flawed core, thereby increasing the total input length. Figure~\ref{fig:pai_vs_paiplus} presents a single-instance case study and should not be interpreted as an aggregate result. The scores returned by both services vary substantially and non-monotonically as the input length changes. For PAI Judge, the average subscore ranges from approximately 3.0 to 8.0, while the overall score ranges from approximately 4.0 to 8.0. Both reach approximately 8.0 at an input length of 2{,}000 characters. PAI Judge Plus exhibits larger fluctuations. Its average subscore reaches nearly 9.0 at approximately 610 characters, whereas its overall score decreases to approximately 1.0 at input lengths of 1{,}005 and 1{,}585 characters. At 2{,}000 characters, the average subscore and overall score increase to approximately 8.5 and 8.0, respectively. These observations show that task-irrelevant suffixes can substantially manipulate the scores assigned to an unchanged flawed response. However, as this experiment is a single-instance case study, broader evaluation is required to determine the prevalence of this behavior.
}

\begin{figure}[t!]
\centering
\begin{subfigure}{0.49\linewidth}
\centering
\includegraphics[width=\linewidth]{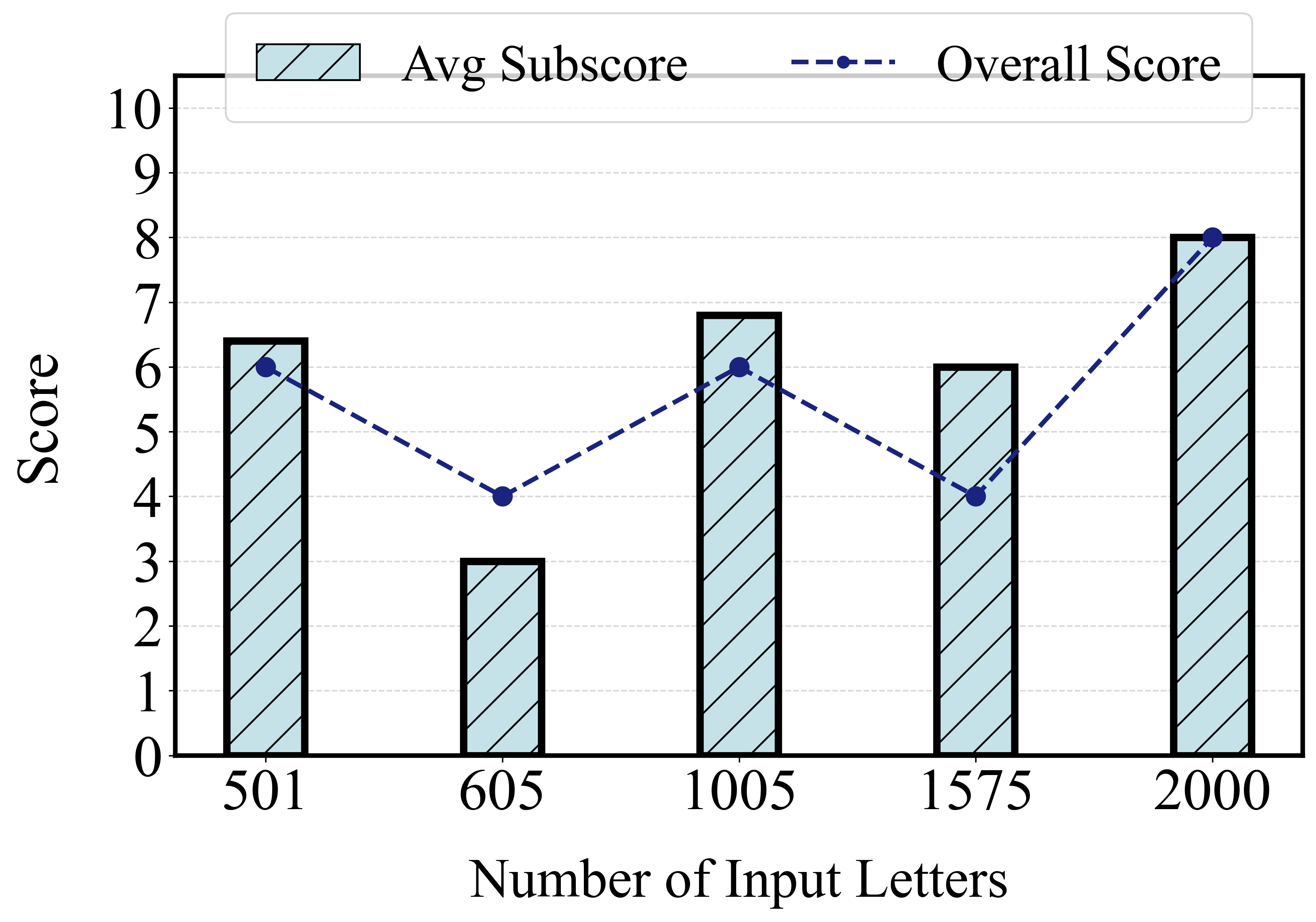}
\caption{PAI-Judge}
\label{fig:pai}
\end{subfigure}
\hfill
\begin{subfigure}{0.49\linewidth}
\centering
\includegraphics[width=\linewidth]{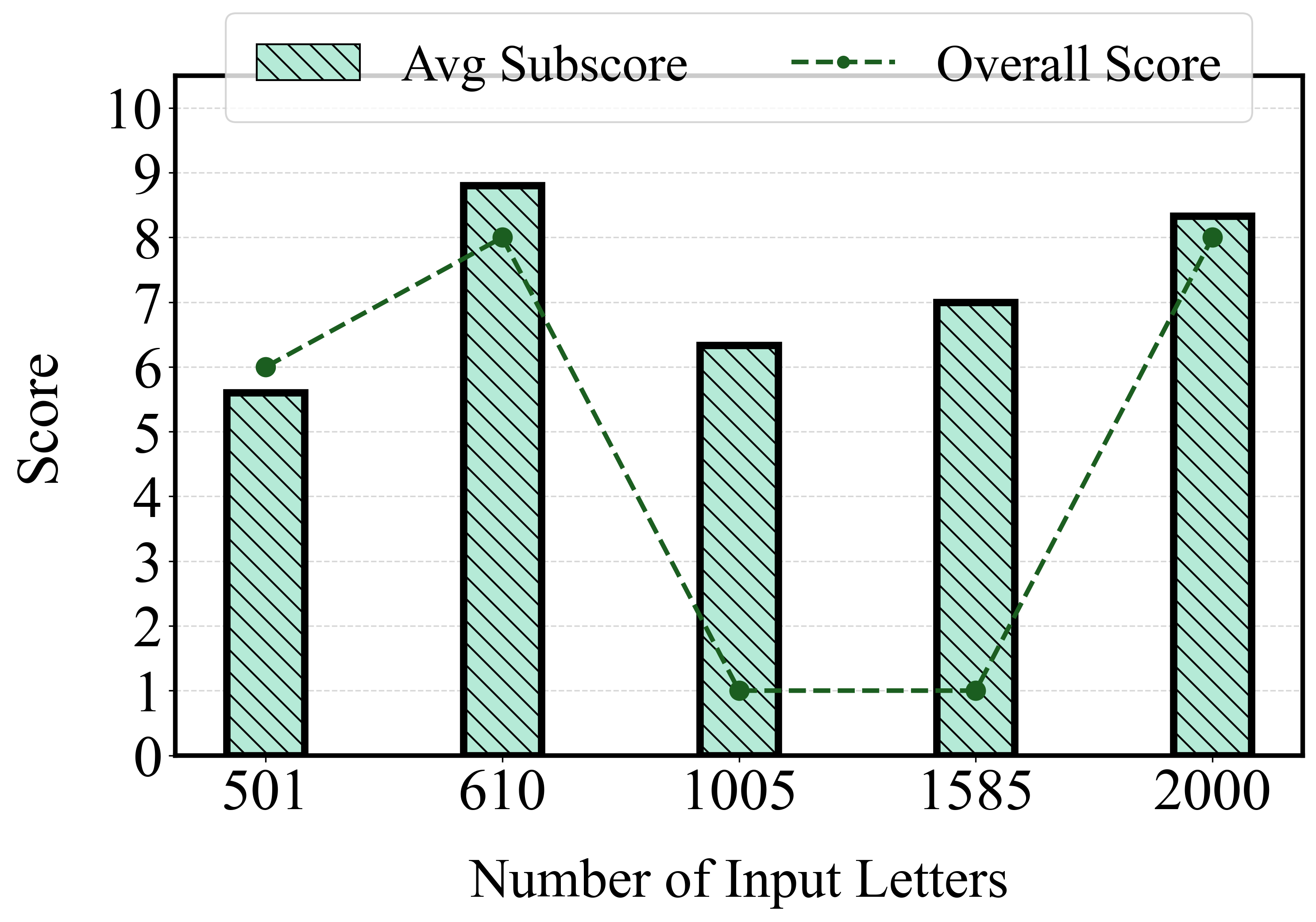}
\caption{PAI-Judge-Plus}
\label{fig:pai_plus}
\end{subfigure}
\caption{{\color{black}Impact of the composite attack on PAI-Judge variants.}}
\label{fig:pai_vs_paiplus}
\end{figure}

\begin{tcolorbox}[
  colback=gray!10,   
  colframe=black!80, 
  boxrule=0.4pt,
  arc=2pt,
  left=2pt,
  right=2pt,
  top=2pt,
  bottom=2pt,
  fontupper=\small,
  before skip=4pt,
  after skip=4pt
]

$\bullet$ \textbf{Finding 8:} We identify a loophole in both PAI-Judge and PAI-Judge-Plus, where combining PAIR-optimized adversarial inputs with long suffixes can compromise the platform's judgment reliability.

\end{tcolorbox}
\section{Limitation}
\subsection{Defense-aware adaptive attacks}
\textcolor{black}{
Our current evaluation does not assume that the attacker has full knowledge of the deployed defense mechanism. A stronger threat model is defense-aware adaptation, where the attacker knows the system prompt, filtering rules, or detection mechanism, and optimizes the attack to bypass them. This setting is especially relevant for deployed systems with fixed defense configurations. We leave defense-aware adaptive attacks as an important direction for future extensions of \sys.}
\subsection{Dynamic multi-turn attacks}
\textcolor{black}{In our setting, a multi-turn conversation can be represented as a fixed dialogue history. This history is constructed before the request is sent, using information available to the attacker. Therefore, our evaluation can support attacks against a fixed conversation context. However, we do not include dynamically staged attacks where the attacker observes intermediate responses and adapts later turns during deployment. This stronger setting requires an interactive evaluation protocol and is left for future work.}

\section{Conclusion}
\label{sec:conclusions}

This work presents the first scalable, automated framework for evaluating LLM-as-a-Judge robustness under adversarial attacks. Our benchmarking shows that current evaluators (even  Alibaba’s PAI platform) remain vulnerable and biased, underscoring the need for stronger defenses and more reliable real-world deployment. 
We also discovered that robustness varies substantially across prompt templates. Our findings provide actionable insights for deploying trustworthy LLM-as-a-Judge systems in security-sensitive applications.

\bibliographystyle{plain}
\small
\bibliography{reference}

\end{document}